\documentclass[a4paper]{cas-dc}

\usepackage[T1]{fontenc}
\usepackage{hyperref}
\usepackage{paralist}
\usepackage{microtype}
\usepackage{acronym}
\usepackage{booktabs}
\usepackage{multirow}

\usepackage{subcaption}
\usepackage{caption}
\usepackage[numbers,sort]{natbib}
\graphicspath{ {./images/} }

\usepackage{listings}
\usepackage{pifont}

\usepackage{pgfplots}
\usetikzlibrary{patterns}
\newlength\figurewidth{}
\newlength\figureheight{}
\pgfplotsset{width=7cm,compat=1.8}

\begin{document}

\shorttitle{Selective Imaging on Live Systems}
 \shortauthors{Faust et al.}

\title[mode = title]{Selective Imaging of File System Data on Live
	Systems}


 \author[fauaddress]{Fabian Faust}
 \ead{fabian.faust@fau.de}
 \author[quosec]{Aur\'{e}lien Thierry}
 \ead{a.thierry@quosec.net}
 \author[fauaddress]{Tilo M{\"u}ller}
 \ead{tilo.mueller@cs.fau.de}
 \author[fauaddress]{Felix Freiling}
 \ead{felix.freiling@cs.fau.de}

 \address[fauaddress]{Friedrich-Alexander-Universit\"at Erlangen-N{\"u}rnberg (FAU), Germany}
 \address[quosec]{QuoSec GmbH, Frankfurt/Main, Germany}


\begin{abstract}
In contrast to the common habit of taking full bitwise copies of storage devices before analysis, selective imaging promises to alleviate the problems created by the increasing capacity of storage devices. Imaging is selective if only selected data objects from an image that were explicitly chosen are included in the copied data. While selective imaging has been defined for post-mortem data acquisition, performing this process \emph{live}, i.e., by using the system that contains the evidence also to execute the imaging software, is less well defined and understood. We present the design and implementation of a new live \emph{Selective Imaging Tool} for Windows, called SIT, which is based on the DFIR ORC framework and uses AFF4 as a container format. We discuss the rationale behind the design of SIT and evaluate its effectiveness.
\end{abstract}


\maketitle


\section{Introduction}

While the overall approach of forensic investigations of storage
devices has changed little over the last decade, the amount of data
that needs to be processed keeps increasing. Digital forensic
investigators are therefore facing growing problems caused by
technological advances in the size of storage
devices. \citet[p.v]{bigDigital} summarize the situation as
\begin{quote}
  ``we are drowning in a deluge of data, more and more every day.''  
\end{quote}
These problems are amplified by the common habit of \emph{forensic
  imaging}, namely to create full 1:1 images of every byte stored on a
device before any further investigation of the storage device is
performed.  Overall, the inadequacies of forensic imaging as
prerequisite to any form of forensic investigation are apparent and
well-known today.

In the literature, \emph{selective imaging}
\cite{SelectiveImagingRevisited} has been advocated as the solution to
these problems: The term refers to the process of only copying
selected data objects, thus creating a \emph{partial image} that needs
considerably less time and space to be taken. Despite multiple
proposed concepts for the classical post-mortem acquisition approach,
however, the range of dedicated selective imaging tools remains
scarce. As \citet{sack} reports from interviews with practitioners,
while forensic investigators do regularly end up recognizing the need
for a selective approach, they often use tools that neither are
intended for forensic investigations nor fulfill even basic
requirements of forensic soundness.

While selective imaging has been discussed in the context of (public)
forensic investigations by law enforcement, it is also of great
relevance to (private) forensics investigations performed by
specialized companies within organizations, mainly with the goal of
confirming the existence of unwanted software, unlawful actions by
employees or external hacking attacks. The primary goal is often to
gather as much relevant evidence about actions taken and caused by
malicious third parties or software in the shortest amount of time
possible. This is accompanied by the requirement that systems cannot
be turned off for imaging, giving rise to \emph{live forensics}, i.e.,
the capture of evidence using the same system to access its data. This
is in stark contrast to the traditional post-mortem acquisition of
data, for example, from a hard drive after shutting down the system
\citep{ITForensikDennisHeinson}.  Especially in larger companies,
software is deployed that allows for a
\emph{triage}~\citep{ForensicTriageAndIncidentResponse}, an initial
classification of people, data, and objects into different priority
categories for later manual analysis. However, generally, such software is not developed or intended for live selective imaging, and therefore not forensically sound for this approach. 

To summarize, there is ample need for selective live acquisition of
file system data within forensic investigations, but there is a
definite lack of concepts and tools for performing this task.

\subsection{Contributions}

In this paper, we present SIT, the \emph{Selective Imaging Tool} that
can perform selective imaging of file system data on live Windows
systems. Through a carefully crafted design, SIT achieves a high
degree of forensic soundness to safeguard the evidential value of the
acquired partial image. SIT is based on modern investigative software
components such as the DFIR ORC framework~\citep{DFIRORCGithub} and
file formats such as AFF4 as its forensic container format. SIT is
fully open-source and available on GitLab. We are not aware of any
other open-source tool that allows the collection of evidence from live
systems with similar degrees of reliability and integrity.

\subsection{Roadmap}

We give a general introduction to the literature of selective imaging in Section~\ref{sec:definition-of-selective-imaging}. We then discuss relevant requirements for the special case of selective imaging for live systems in Section~\ref{sec:selective-imaging-on-live-systems}. We then present SIT in Section~\ref{sec:implementation} and its evaluation in Section~\ref{sec:evaluation}. We conclude in Section~\ref{sec:conclusions}.

\section{A Brief History of Selective Imaging}
\label{sec:definition-of-selective-imaging}

The concept of selective imaging goes back to \citet{UnificationOfDigitalEvidenceFromDisparateSources} and his idea of a \emph{Digital Evidence Bag} (DEB), a universal container for the capture of arbitrary and arbitrarily fragmented digital evidence. In wise foresight, \citet{UnificationOfDigitalEvidenceFromDisparateSources} did not only design DEBs for classical post-mortem storage captures but also as a flexible concepts for the capture in ``real-time'' and ``live'' system environments \cite{UnificationOfDigitalEvidenceFromDisparateSources}. Each of these bags contained the collected data objects, their associated metadata, and the DEB-specific metadata for localization, identification, and integrity assurance.

Full 1:1 imaging operates on the device level, with each device being an
indivisible object that can be either imaged completely or not at
all. The idea of \emph{selective imaging}
\cite{SelectiveImagingRevisited} is performed on the higher levels of
abstraction in the storage hierarchy \cite{Carrier}, with the file
system level being the initial choice for most cases. Since
actions performed on one level have implicit consequences to lower
levels, they can only be fully captured if these levels are also
included in the process \citep{StuettgenSelectiveImaging}. If an
artifact is selected to be included in the image, the stored data
should contain the artifact itself and all the metadata associated
with it, as well as the metadata required to uniquely identify and
locate the source data object, at the time and state of the system on
acquisition. This is the basis for the formal concept of a
\emph{partial image} \cite{SelectiveImagingRevisited}.  This also
allowed to collect data objects of different levels of abstraction at
the same time, be it a file, a record within a file, or an unallocated
sector on the physical layer of a storage device.

Interviews conducted by \citet{StuettgenSelectiveImaging} amongst
forensic investigators, both law enforcement and private, illustrated
the widespread usage of standard copying tools like \emph{Microsoft
  Windows Explorer} and \emph{X-Copy} to selectively copy relevant
files. The results also showed findings similar to a survey performed
by \citet{sack} in that, while a majority of the forensic
investigators have performed selective imaging at least once in a
case, the confidence regarding the acceptance of acquired evidence in
court is less established. Potential concerns raised by legal scholars
and practitioners refer to the \emph{completeness}, the
\emph{reliability}, the \emph{integrity} and the possibility of
missing evidence in \emph{slack} or \emph{unallocated space} of a
selective image.

Live analysis refers to forensic analysis performed on a live system,
i.e., using the hardware and software of the system to be
investigated. Similarly, live selective imaging is selective imaging
on live systems. Although of eminent practical interest, we are not
aware of any systematic treatment of this topic in the literature, let
alone the existence of a tool that can perform live selective imaging
in a forensically sound manner. This includes tools that offer live forensic functionality such as \emph{EnCase Forensic Imager} \citep{EnCaseForensicImager} and \emph{FTK} \citep{ForensicToolKit}, as researched by \citet{sack}. There is, however, much literature on live
analysis in the context of memory forensics, a topic with many but yet
uncharted similarities to live file system imaging: In memory
forensics, RAM of a running system is acquired, giving rise to
challenges of forensic soundness \cite{DBLP:journals/di/VomelF12}, and
risks of anti-forensic software and data corruption
\citep{MemoryForensicsPathForward}.

\section{Selective Imaging on Live Systems}
\label{sec:selective-imaging-on-live-systems}

Despite the fact that a live selective imaging approach offers a variety of significant benefits and can be the only option in certain cases, the challenges caused by the nature of working on live systems, that are often beyond the investigator's control prior to access, are significant. Some of these challenges make problems inherent to selective imaging in general more critical, while others originate solely from the live environment and the options for anti-forensic interferences with the investigation it offers. Before we develop general criteria for selective imaging on live systems, we briefly revisit the concept of forensic soundness from the literature.

\subsection{Forensic Soundness}

\emph{Forensic soundness} serves the goal of ensuring that the collected evidence is not altered in any way from the source. \citet [p.8]{WhenIsDigitalEvidenceForensicallySound} defines forensic soundness as ``the application of a transparent digital forensic process that preserves the original meaning of the data for production in a court of law.''

According to \citet{WhenIsDigitalEvidenceForensicallySound} the main priorities of forensic soundness can be summarized as follows:
\begin{enumerate}
	\item The acquisition and subsequent analysis of electronic data has been
	undertaken with all due regard to preserving the data in the state
	in which it was first discovered.
	
	\item The forensic process does not in any way diminish the evidentiary
	value of the electronic data through technical, procedural or interpretive
	errors.	
\end{enumerate}

Naturally, these rather abstract requirements have to be interpreted
in the legal framework in which evidence is processed. Such standards
were formulated by \citet{Safeguarding} in the context of the forensic
analysis of cryptocurrencies, thus generalizing the notions of
\citet{WhenIsDigitalEvidenceForensicallySound}:
\begin{itemize}

\item The processing of the data must be compliant with the legal
  framework it takes place in (\emph{lawfulness of data processing}).

\item The \emph{authenticity and integrity} of data must be ensured in
  such a way that allows an assessment of the evidential value in
  trials (chain of custody). If data is changed, it must be clear how
  the alteration exactly changed the data.
  
\item The processing of data must be \emph{reliable}, i.e., based on
  scientific verifiability and testing.

\item Investigators using specific techniques must be qualified to use
  them (\emph{qualification}).

\item The method for collecting data and gaining information must be
  repeatable and reproducible (\emph{verifiability}).
  
\item Conclusions drawn from the evidence must be logical, consistent
  and compelling (\emph{chain of evidence}).

\item Concerned parties have the right to inspect records
  (\emph{disclosure of evidence}). In contradictory criminal
  procedural systems (such as the one used in the US), this implies
  the right to disclose case-relevant evidence by the public
  prosecutor's office. In inquisitorial criminal procedural systems
  (such as Germany's criminal procedural law) this is realized as the
  right for the accused to inspect the evidence gathered by the
  police.

\end{itemize}
We will apply these requirements to the live acquisition of file
system data shortly.

\subsection{Forensic Soundness on Live Systems}
\label{sec:forensic-soundness-on-live-systems}

While it has been pointed out \citep{DigitalEvidenceAndComputerCrime}
that even the routine task of post-mortem data acquisition from a hard
drive with a write-blocker alters the original state of the source,
the copied bits of data are usually acquired in a reliable and
repeatable manner since many influencing factors like the hardware and
software used for copying are under full control of the analyst. The
situation for live systems is totally different since the used
hardware might fail during acquisition and the operating system of the
target system might have been manipulated in diverse ways. The
situation is therefore similar to the acquisition of volatile evidence
such as RAM \citep{DBLP:journals/di/VomelF11} or data about
cryptocurrencies from a blockchain \citet{Safeguarding}. The general
requirements also have to meet the evidence standards of the legal
system in question.

Reconsidering the requirements of \citet{Safeguarding}, the main
problems relevant to forensic soundness on live systems are the
\emph{verifiability} of the data collection methods and the problem of
maintaining \emph{authenticity and integrity} at all times.

\subsubsection{Maintaining Authenticity and Integrity}

Live systems are moving targets on which data is continuously changing. Furthermore, the use of any analysis software may have side-effects on the system itself. This can be reduced by decreasing resource usage as much as possible, by executing the software from an external flash drive and avoiding write operations on the system storage. Therefore, avoiding the usage of any system storage for temporary files is important and can be replaced by creating a custom temporary directory on the flash drive. 

Interference with other (wanted or unwanted) software running on the system which may have an effect on the system's data and possible sources of evidence cannot always be prevented or reasonably predicted. Software should therefore use access methods that create the least interference with the running system, and after acquiring any artifact the acquisition method should use integrity protection techniques (such as cryptographic hashes) to prevent unnoticed \emph{a posteriori} manipulation, storing them and any associated metadata reliably and securely. In order to further prevent interference, the time spent executing software on the target system should be kept to an absolute minimum. To facilitate this, software used has to be developed with a priority on efficiency and minimal resource usage, to keep corruption limited.

\subsubsection{Verifiability}

In post-mortem data acquisition, the results could usually be
independently verified by turning back to the original.  In live
analysis, technical circumstances often prevent the repeatability or
reproducibility of data acquisition, considerably decreasing its
evidential value \citep{Safeguarding}. It is therefore important to at
least make the acquisition process as plausible as possible in an
\emph{a posteriori} context.

Two methods can help to achieve this. The first is robustness and extensive error handling of the tool to unforeseen conditions like hardware or software failure and unexpected shutdowns. The error handling priority should always be
on preserving the collected artifacts and their integrity. As repeatability at a later time may not be possible or useful, a validation step following the acquisition phase would be advisable to check the collected evidence for obvious corruptions or missing data. This allows for immediate reaction without the need to do a manual analysis, for example by changing acquisition parameters and repeating the process.

The second method to achieve plausibility of the acquisition process
is extensive log documentation. Key steps in the imaging process also
need to be communicated to the user to further facilitate a quick
reaction. For example, in case of unexpectedly long imaging duration
or imminent hardware failure, the investigators may decide to stop the
process and prioritize other data samples. All data acquired up to
this point should still be reliably stored and documented.

\subsection{Concrete Requirements}\label{sec:rules} 

Based on the above discussion, we now derive a set of concrete requirements for
selective imaging with the goal to maximize forensic soundness on live systems. We concentrate here on the acquisition process. Since live selective imaging may have to be performed outside a secure and monitored laboratory, once the evidence is collected,  measures must be taken to ensure the evidence is in turn secured and monitored from the moment it is physically extracted from the live system, ideally on a removable flash drive, up to the moment it is delivered into a suitable laboratory. 

The following rules are categorized into five main priorities, with each priority serving one of two main objectives. Preserve the acquired data in its original meaning as much as possible and allow independent evaluation of the entire process without source access:  

\begin{itemize}

\item Minimize source corruption
  \begin{itemize}
  \item Minimize side effects of used software
  \item Minimize operation time on the live system
  \end{itemize}
	
\item Ensure evidence data authenticity and integrity
  \begin{itemize}
  \item Collected data must not be changed from the source version 
  \item Calculate at least two different hash codes upon acquisition 
  \item Verify evidence data integrity using hash codes
  \end{itemize}
	
\item Provide extensive documentation
  \begin{itemize}
  \item Every step taken must be documented 
  \item Key steps must be communicated to the user
  \end{itemize}
  
\item Ensure digital reliability and security
  \begin{itemize}
  \item Software used for investigation must be developed with a focus
    on reliability and security
  \item Measures against attacks and interferences by third party
    software must be present
  \item Collected data must be stored in a reliable and secure digital
    format
  \end{itemize}
  
\item Ensure physical reliability and security
  \begin{itemize}
  \item If performed outside a forensic laboratory, the collected
    evidence must be secured and monitored, until the delivery of the
    evidence into a secure and monitored laboratory
  \end{itemize}
  
\end{itemize}

\subsection{Documentation of the Selection Process}

The selection process itself, whilst being an integral part of the
selective imaging concept, should on an implementation level be
treated separately from the actual imaging tool containing all the
features required to proceed after the selection targets have been
chosen. This is due to the fact that for the selection itself, any
analysis tools that have minimal side-effects on the machine may be
used. This includes both statically pre-selected lists of files to be
acquired as well as a live analysis involving manual browsing and file
selection. To satisfy the aspect of \emph{verifiability} and the
\emph{chain of evidence}, the selection process must be sufficiently
understandable, increasing the burden of documentation in case live
browsing and manual selection are chosen.

\section{SIT Design and Implementation}
\label{sec:implementation}

 The main functionality of SIT is to allow the selective collection of forensic artifacts on file system level, alongside key metadata, validating the results to detect unexpected results and external interferences, integrating the results into an AFF4 forensic image, and then verifying the artifacts using hash codes, all while maintaining the new live forensic soundness rules. The project is available in its GitLab repository \citep{SITGithub}.

\subsection{Design \& Architecture}

The main goal of our implementation was to create a modular framework for selective imaging on live Windows systems that implements the rules established in Sect.~\ref{sec:rules}. In order to achieve this, the software had to be portable, i.e. a binary that can be moved between different systems without the need for a prior installation and which is running with minimal external dependencies. Furthermore, in addition to the execution from an external flash drive, the usage of a custom temporary directory, and the extensive verification using more than one hash code for each artifact, more secondary measures were implemented.

A separate validation step after the artifact acquisition phase creates a redundancy to identify obvious interferences and attacks on the acquired data by malicious software or corruption caused by errors. As a suitable storage container, the \texttt{AFF4} format was chosen for its direct artifact-metadata association mechanic, intuitive metadata representation using RDF turtles, and the lightweight compression algorithm \citep{ExtendingAFF4}. Providing extensive user feedback and logging was another priority, alongside sufficient error handling for basic security. Since external libraries need to be statically-linked in order to maintain portability and compatibility with different Windows versions, an aspect which increases the software's footprint and decreases its efficiency, one secondary goal was to directly implement simpler functions such as RDF serialization, thus avoiding the usage of a library. Lastly, a backup archive of all acquired artifacts serves as another redundancy in the case of data corruption.   

As a foundation, SIT is using the DFIR ORC framework \citep{DFIRORCGithub} to create a single portable preconfigured binary that can be run as a command-line tool. The intention is to execute it from an external flash drive in order to prevent overwriting files on the system storage devices and to have an option to extract the results.
DFIR ORC, short for \emph{Digital Forensics and Incident Response,  Outil de Recherche de Compromission} is a framework for digital forensic tools on live systems and comes with a collection of specialized tools for different types of forensic artifacts. It is developed by ANSSI, the National Cybersecurity Agency of France, and is at the release of this work still being actively updated \citep{DFIRORCdocumentation}.

As forensic soundness is dependent on the tools that are being used, the DFIR ORC framework itself supports this in multiple categories. Firstly minimizing its footprint on the system, the output is stored in an archive that is constantly updated during the execution to secure the results and minimize the use of temporary files. Secondly, it allows scheduling tools with bigger impact on the system last and performing other more lightweight tools first. Thirdly it ensures that data integrity is maintained by storing the collected artifacts as soon as possible and computing hash values on acquisition to allow verification of the data integrity at any time \citep{DFIRORCdocumentationDesignPrinciples}. 

The creation of a single portable binary is performed by what is called the configuration process. In this step, a script is run to combine two compiled binaries containing the DFIR ORC framework code, any number of custom binary tools, and a set of configuration XML-files, by executing the integrated \emph{ToolEmbed} software. The DFIR ORC binaries serve as a mothership or base for the creation and execution of the configured binary. It provides the execution framework and includes both the first code that is executed and a pre-embedded suite of forensic tools.

One of the advantages of DFIR ORC is that it was developed with efficiency, as well as reliability and security in mind. For example, it can look up and use resources from its internal parent and grandparent processes without the need for unnecessary file extraction. In addition to the constantly updated output archive, it also has extensive logging and error handling features \citep{DFIRORCdocumentationArchitecture}.

\begin{figure*}
    \centering
    \includegraphics[width=0.7\textwidth]{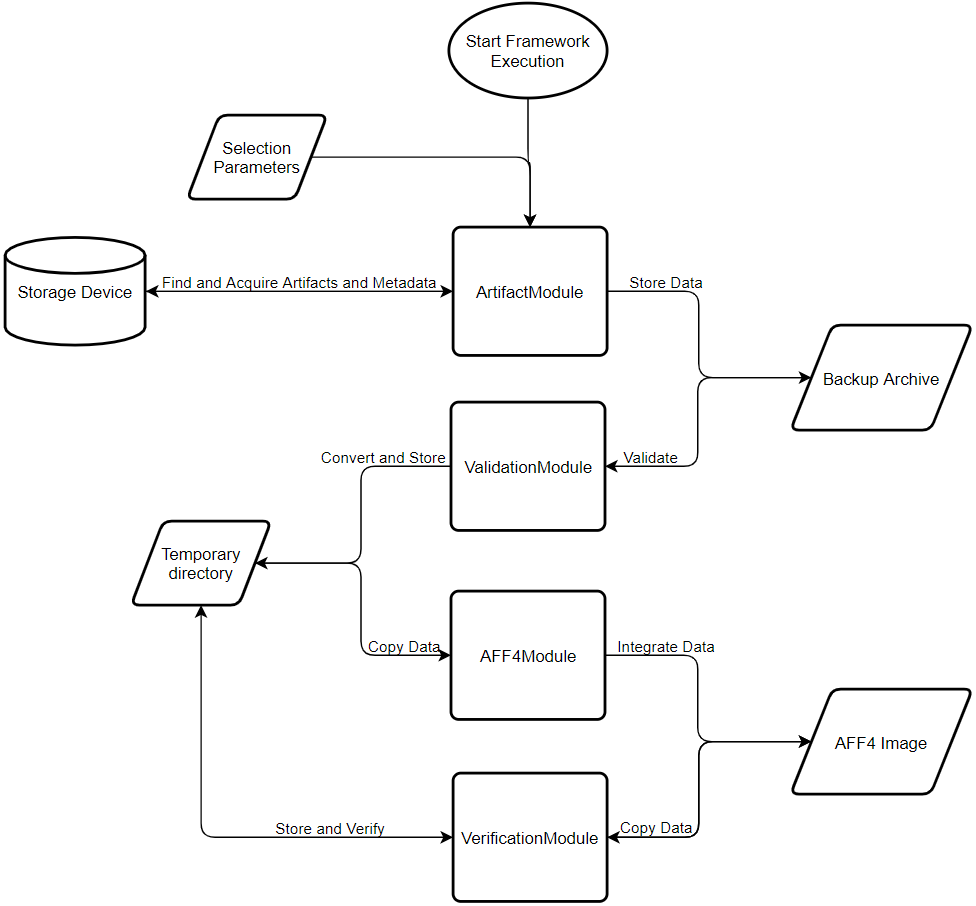}
    \caption{Overview of the SIT architecture and its modules.}
    \label{fig:SITModules}
\end{figure*}

\subsection{SIT Modules}

SIT is made up of four logical modules. They are designed to run sequentially, as illustrated by Figure \ref{fig:SITModules}, and while each module is built to work with the results of the previous one, they can also be repeated, executed independently, or disabled. If efficiency is a priority, disabling modules such as the Verification Module will improve performance at the cost of an on-site integrity check. In case of an unexpected shutdown by the live system or crashes during the imaging process, the intermediate results up to that point can be used to continue the process. In addition to the SIT modules, any external binary tool can be integrated into the portable binary.

Each module is giving extensive user feedback via console output and creates logs of every relevant action taken. The console feedback allows the user to react to unexpected behavior by the software, unusually long acquisition times, or anticipated system failure on longer operations, by stopping the process at a suitable process step, while retaining the results up to that point. It is then possible to restart the process with different parameters, including previously disabled modules.

\paragraph{Artifact Module}

The artifact module consists of a modified version of a DFIR ORC tool called \emph{GetThis}, developed as a forensically sound all-purpose file acquisition tool. It can acquire files by searching file system entries for parameters such as name, path, and size, determined during the configuration process. Its main focus is on NTFS file system entries, as for modern Windows systems, XP and newer, this is the default file system, with alternatives such as FAT being more relevant in external storage devices for example. Each file artifact is copied without changing the source and a wide range of metadata categories is also acquired, including MD5, SHA1, and SHA256 hash codes created immediately after acquisition. The results are then stored in ZIP archives, serving as backup and base for the next steps. Inside the archive, the metadata is temporarily stored using a CSV file. 

\paragraph{Validation Module}

Once the artifacts and metadata are acquired and stored in the backup archive, the validation module is executed, serving as a fail-safe that aims to identify unusual results and inform the user. The artifact module interacts most with the file system and files stored on the system and is therefore very vulnerable to interferences, data corruption, and crashes. As the goal is to detect such cases as quickly as possible, three steps are involved. Firstly each module is responsible for handling its errors reliably and documenting any unexpected behavior. Secondly, the artifact and metadata output is validated by the validation module checking if any inconsistencies, such as missing metadata for collected artifacts, missing artifacts for collected metadata, or incorrect data types can be identified. Lastly, verification of data integrity is done in the last step, the verification module. In addition to validating the output from the artifact module, the validation module also converts the metadata into an RDF turtle, in preparation for integrating it into the central metadata registry of the AFF4 image. 

\paragraph{AFF4 Module}

As described in Sect.~\ref{sec:definition-of-selective-imaging}, suitable storage formats for forensics should have certain features. \emph{AFF4}, short for Advanced Forensics File (Format) 4, is an open, ZIP-based, extensible file format for storing evidence and case-related information. It uses an object-oriented approach to store data objects and metadata, using a central data store, called the resolver to manage references between objects. Each reference is maintained using Uniform Resource Identifiers (URI), made up of either internal AFF object Uniform Resource Names (URN), uniquely generated as part of the aff4 namespace, or a more general Uniform Resource Locator (URL). Every AFF4 object has its own URN and can therefore be internally identified by the resolver and associated with its metadata \citep{ExtendingAFF4}.

Metadata is a central aspect of AFF4 and can also exist independently from a data object. It is stored in (Subject, Attribute, Value) tuples inside a central RDF turtle file, bundled into unique URN entries, which allow $direct$ $association$ with the corresponding file object or identification as an abstract metadata object. Metadata that is not part of the AFF4 created data such as compression or size, is internally stored using an XML Schema Definition (XSD) type such as xsd:string or xsd:dateTime \citep{HashBasedDiskImagingUsingAFF4}. While each object has its own URN, URLs may be used interchangeably with a URN to facilitate the sharing of evidence files between investigators \citep{ExtendingAFF4}.   

\paragraph{Verification Module}

The verification module is part of the AFF4 module's source code, to improve efficiency and reduce storage space of the SIT binary, as it uses the same functions to access the AFF4 image. To perform the hash verification, all the artifacts stored in the AFF4 image are copied into the temporary directory and the MD5, SHA1, and SHA256 hash codes are calculated. These are then compared to the hash values collected by the artifact module upon acquisition and stored in each artifact's metadata set. Should any artifact have a missing hash code entry, the code calculation fail for any reason, or the codes not be equal, the verification is considered not successful for this artifact and the user is notified. The usage of multiple hash codes is intended to make unnoticed attacks on the integrity of collected evidence more difficult and in the case of SHA256 provide a collision resistant verification option conforming to current NIST \citep{NISTHashPolicy} guidelines.      

\section{Evaluation}
\label{sec:evaluation}

SIT includes a variety of different measures implementing the rules established in Section \ref{sec:rules}. Their effectiveness was evaluated according to the corresponding priorities and rules. As 100\% effectiveness is not realistically possible, due to the wide range of factors that are impossible to predict and control, and since there is no reasonable way to quantify the forensic soundness level, especially in live forensics, the goal was to reach a level of forensic soundness comparable to physical forensic investigations. This represents the balance between having to lower the forensic standard below complete perfection on one side, as attempted by full 1:1 imaging inside secure laboratories, and the requirement to maintain a sufficiently high level of forensic soundness to not cause justifiable doubts in court about the evidence collected in this manner.

As such the main question that was answered for each measure to be evaluated for effectiveness was if concrete doubts about the forensic soundness were justifiable. This was done separately for each of the rules except for the rule to ``ensure physical reliability and security'', which has to be maintained in a non-digital environment and exceeds the scope of this work. 

\subsection{Minimizing Source Corruption}\label{sec:minimizing-source-corruption}

The rules for achieving the goal of minimal source corruption were to minimize both the side effects of the used software, as well as the operation time spent on the live system. 

The concrete measures taken to achieve this were to use an external flash drive to execute the tool from and store the collected evidence on, as well as to avoid writing data on the system storage drives by using a custom temporary directory on the flash drive. Additionally, SIT was developed with maximum efficiency in mind, which includes the option to easily disable any module to improve the execution time and therefore reduce source corruption, as well as the option to limit the memory and time usage. 

To evaluate the actual effect these measures had, it is necessary to divide the areas of possible source corruption. 

One is the RAM of the live system, volatile memory that loses all its content once the system is shut down. Overwriting data in this memory is therefore mainly problematic if the system has not been shut down since the last time a potential suspect has had access or relevant operations have been performed using the system. In that case, executing the tool might overwrite data stored in this memory, depending on the capacity of the RAM, the remaining free size, the memory demand fluctuations due to other currently active software, and whether the data stored there is at risk of suddenly being freed. Due to the large amount of unpredictable and uncontrollable factors, the most realistic approach for minimizing the corruption of data stored in the RAM, is to decrease the amount of space the used software requires there. This is primarily done by prioritizing this aspect when developing such a tool, which in the case of SIT includes actions like storing acquired artifacts in the target archive as soon as possible to remove them from the memory or reducing the usage of external libraries whenever feasible, because portable software requires the static inclusion of these libraries, increasing the memory usage. The main advantage however is that during the configuration process an upper limit for the memory usage can be set to any value. While this can lower the execution speed, it allows the investigator to prioritize RAM data integrity to any degree desirable. For this reason, the memory integrity can be maintained to a limited degree concerning corruption caused by SIT, however due to the changes that are likely to be caused by other software, fully maintaining memory source integrity is not possible on a live system. 

Avoiding source corruption on system storage devices is considerably easier than for RAM, as software is not required to store or modify data on them and can instead use external drives such as an external flash drive. Retrieving files without inducing updates to their respective timestamps (and changes to the MFT) is performed through the underlying DFIR ORC framework. DFIR ORC interacts directly with the volumes to parse the MFT and NTFS data without using the operating system's specific system calls. Additionally, SIT uses a temporary directory on the flash drive to store its files whenever feasible. If done consequently and no other changes are directly initiated on data from the storage device, this reduces the source corruption on the system storage devices to actions performed by other software in general or in response to the tool's execution. Any software active on the system may store or modify data as a reaction to SIT's execution, but the most common source is the operating system, which constantly manages active software and may create log files or change configuration files for any reason. These constant changes to logs or operating system files should be considered when they are targets for acquisition, however they have a minimal likelihood to create source corruption large enough to change relevant evidence data and therefore when used in conjunction with consistent usage of custom external temporary directories, the risk is in most situations limited. However a significant uncontrollable factor is anti-forensic software that deliberately causes source corruption to obfuscate or remove evidence, possibly as soon as it notices external factors such as a live forensics tool. 

\medskip 

Due to this factor and the expected RAM corruption, despite all measures taken, for live system forensics the likelihood of source corruption remains significant. The question that needs to be answered is if this risk is high enough to justify significant doubts in the evidence collected on this system using live selective imaging.

To give a possible answer to this question one needs to consider the physical counterpart of crime scenes. It is possible for a suspect to destroy the evidence or manipulate it prior to or on arrival of the forensic investigators. This can be done for example by laying a fire or adding false trails. In many situations there is no realistic way to prevent this from happening prior to arriving on the scene, so the best forensic investigators can do is to investigate if any such action was performed and find resulting evidence. Still, the risk of missing evidence that used to be there would be significant, but in absence of better alternatives, out of necessity for a solution, and due to the low likelihood of this happening, evidence collected on such a scene is usually considered admissible in court. 

In a similar manner, if such an action to destroy, corrupt or obfuscate evidence on a live system is evident, further investigation on this action could be sufficient to have the collected evidence from this system be considered admissible in court. During the investigation, it would then be necessary to evaluate how likely it is that the evidence was corrupted. 

\subsection{Ensuring Evidence Data Integrity}

Ensuring evidence \emph{data integrity} entails three rules. Collected evidentiary data must not be changed from the source version, at least two different hash codes must be calculated upon acquisition and data integrity must be verified using these codes.

These rules are implemented in SIT by calculating three hash codes, MD5, SHA1, and SHA256 immediately upon acquisition and storing them alongside the artifact. The integrity of all artifacts is then verified as the last step, the verification module. While the corruption of collected data can not be reliably prevented on the live system, as even encrypted data can be changed haphazardly, it is unlikely to go unnoticed. For this, all hash codes would need to be changed to match the artifact's new data or the SHA256 code would have to be manipulated in secret to match an inserted fake-artifact that is causing a collision with the other two codes. If this is not the case, as soon as one hash verification is not successful, checking other artifact verifications gives additional vital information on whether a random attack or error has caused untargeted data corruptions, or if a pinpoint attack has taken place. Depending on the result, the entire acquisition may have to be repeated with different parameters or the backup archive used to verify the integrity of its contents. 

As it is possible to reliably determine whether collected evidence has been corrupted or not, there should be no justifiable doubts about the forensic soundness of evidence that has been successfully verified. Corrupted evidence however, may have to be collected again, replaced from the backup archive or discarded.

\subsection{Providing Extensive Documentation}

The live forensic soundness rules determined that every taken step must be documented and key steps must be communicated to the user. For SIT the two evaluation criteria are if the documentation presents sufficient information about the live selective imaging process to allow insight for an external investigator and if the user feedback is enough to allow the user to react to the status of the process based on the current progress.

The first criterion can be narrowed down to whether each log file achieves its goal, which is that just from the log file, it becomes possible to determine the actions that were performed by the software or module. When determining the documentation granularity, the balance between quick access to the relevant steps and performance needs to be taken into account. If a log is too superficial it might not provide the necessary information to identify what went wrong and where, but if it is too extensive, it will take too long to navigate and use, especially in potentially time-critical situations such as live forensics. 

Concerning the second criterion, the user needs to be aware of at least the key steps the tool is currently performing in order to make an informed decision on whether to let the software finish its execution or stop it to choose a different action or configuration. 

\begin{figure*}
	\centering
	\includegraphics[scale=0.55]{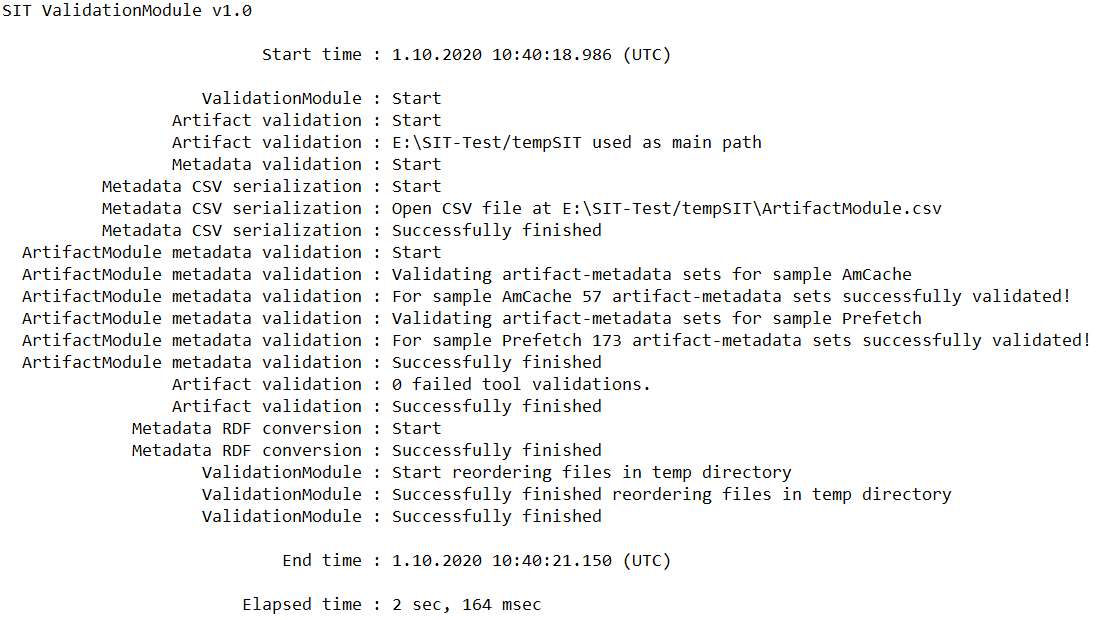}
	\caption{Example log file from the Validation Module.}
	\label{fig:SITExampleLog}
\end{figure*}

\begin{figure*}
	\centering
	\includegraphics[scale=0.55]{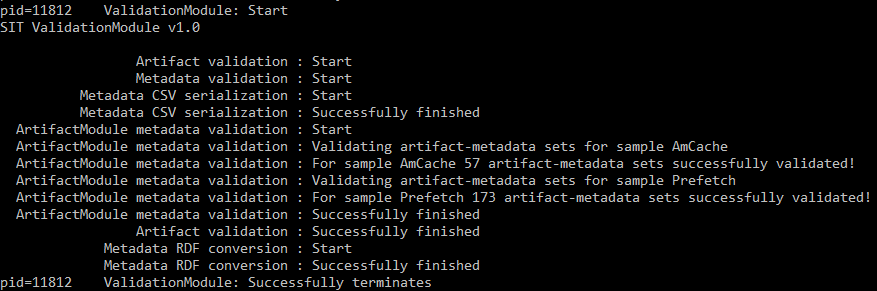}
	\caption{Example output from the Validation Module.}
    \label{fig:SITExampleOutput}	
\end{figure*}

As a result of the extensive documentation, with separate log files created by each module, exemplified in Figure \ref{fig:SITExampleLog}, for SIT an external investigator will be able to determine which actions were performed and if they were successful or have failed, only by reviewing these logs. Due to the console output by each module, as visible in Figure \ref{fig:SITExampleOutput}, the user will have the required knowledge to interact with the imaging process by stopping it at a suitable level of progress, if required.

\subsection{Ensuring Digital Reliability and Security}

The rules for ensuring digital reliability and security required that the software be developed with a focus on reliability and security, measures against attacks and interferences by third party software be present, and the collected data to be stored in a reliable and secure digital format. 

The focus on reliability and security during the development of SIT resulted in the modular structure of the tool and the choice of DFIR ORC as a framework because it offered an already established reliable and secure platform to launch the tool from. The modular structure and resulting compartmentalization allow each module to potentially crash, without affecting the execution of the other modules. If a module requires the output from a previous one to continue, the process can be stopped and restarted from the crashed module. The backup archives also increase reliability by adding a save state to continue execution. As SIT is a portable tool with external dependencies statically included in the binary, the likelihood of errors or crashes being caused due to problems with external dependencies, is avoided. The goal for SIT was to safely stop a module if an error occurred that could not be fixed and relay sufficient information to the user using console output and log files. A further factor for ensuring security was to minimize the risk of abusable weak points in the tool. For this purpose, an effort was made to use secure functions in the code and avoid insecure interactions with user input. 
The validation module acting as a separate redundancy step to check all collected artifacts and metadata for obvious inconsistencies, as well as all evidence data integrity oriented steps offer additional fail-safe functionality. Lastly, the AFF4 format serves the purpose of a suitable storage format, by providing direct artifact-metadata association and therefore quick means to identify missing or corrupted data, as well as facilitating efficient verification by providing reliable access to all metadata.

As a result of the steps taken to ensure digital reliability and security, the risk of attacks and interferences on the tool and its output has been reduced, and SIT should perform as reliably as realistically achievable on an unpredictable live system in general and in case of crashes and errors, appropriate reactive measures are in place to provide information to the user and continue execution as soon as possible.

\subsection{Performance}
When using full 1:1 imaging, an estimation of the total time required for the entire process is realistic as the size of the hard drive is known. The hardware specifications of the system that is used are usually also available. In contrast, artifact samples collected by SIT can potentially be of any size, only limited by the storage capacity of the system drive. While it is possible to identify the size of a sample in advance during the selection process, doing so would require time, effectively increasing the overall duration of the selective imaging process and causing additional source corruption. However, if the size of the sample is not known, the time required can not be estimated in advance and may take significantly longer than expected. Additionally, the hardware of the system that is used may be slow or damaged, while the operating system can be inhibited by a lack of maintenance or software using up resources. Taking this into account, as the user can choose an upper limit for both memory usage and total elapsed time when configuring the binary, it is possible to optimize the performance depending on the current priorities. 

As an example, executing SIT on a new up-to-date Windows 10 computer with modern hardware required about 1,5 minutes, acquiring about 200 artifacts of 30 MB in total, while using the same parameters to gather the exact same artifacts, on a very old system running a badly maintained Windows 7 version with slow hardware ended up taking 10 minutes in total. 

Therefore conclusive statements about the performance of SIT are unrealistic, due to the large number of factors involved, that can neither be controlled nor predicted. 

\section{Conclusion}
\label{sec:conclusions}

Considering the problems that current digital forensic investigations face, such as continuous growth of data pools to investigate, time-critical cases, and systems with limited legal and physical access, approaches other than full 1:1 imaging are necessary. Selective imaging as a possible alternative, especially if performed on a live system, bears a wide variety of serious challenges and problems. Maintaining complete source integrity is not possible and even collected evidence can be corrupted. Anti-forensic tools may remove, obfuscate or hide critical evidence with an unpredictable level of freedom, while independent reviews of the process may have to rely entirely on the provided documentation if the source has been corrupted or is no longer accessible.

Since triage and selective imaging on live systems are nevertheless part of many investigators' toolset during private investigations, it is crucial to evaluate and mitigate their shortcomings.
For this purpose, we presented an adapted set of rules to maintain forensic soundness on live systems. While it is not possible to completely eliminate the various problematic aspects, as we also demonstrated by the implementation of SIT, it is well feasible to achieve an acceptable level of forensic soundness, given the limitations of operating on a live system.

Taking into account the complexity and unpredictability of live environments in digital forensic investigations, there is considerable potential for future work improving and creating rules for maintaining forensic soundness on live systems. Especially when considering software security, the constantly evolving anti-forensics toolkit requires in-depth countermeasures to achieve a sufficient level of reliability and security. Additional work specifically on the initial selection process could also improve the entire approach significantly. 

The SIT implementation could be further enhanced by adding the option to encrypt the collected evidence in order to prevent targeted, external manipulation. The security provided by hash codes could be improved by storing them separately from the image, for example in a picture or E-Mail, therefore making it more difficult to manipulate them alongside the evidence. An evaluation of the forensic soundness achieved by different available live forensic tools such as \emph{EnCase Forensic Imager} \citep{EnCaseForensicImager} and \emph{FTK} \citep{ForensicToolKit} in comparison to SIT would also give further insight into different strategies to handle this issue. Last but not least, different selection tools and selection strategies could be evaluated regarding their effects on the chances of finding relevant data.

\def\UrlBreaks{\do\/\do-}
\bibliography{mybibfile}

\end{document}